\documentclass[prb,aps,groupedaddress,amsmath,twocolumn]{revtex4}

\usepackage[dvips]{graphicx}
\usepackage{dcolumn}
\usepackage{bm}
\usepackage{epsfig}

\begin{document}


\title{Ground state properties of the one dimensional Coulomb gas.}

\author{Michele Casula,$^{1}$ Sandro Sorella,$^{2,3}$ and Gaetano
  Senatore$^{2,4}$} 

\affiliation{
$^1$ Department of Physics, University of Illinois at Urbana-Champaign,
  1110 W. Green St, Urbana, IL 61801, USA\\
$^2$ INFM Democritos National Simulation Center, Trieste, Italy \\
$^3$ International School for Advanced Studies (SISSA) Via Beirut 2,4
  34014 Trieste , Italy \\
$^4$ Dipartimento di Fisica Teorica dell' Universit\`a di Trieste, 
Strada Costiera 11, 34014 Trieste, Italy
} 

\date{\today}

\begin{abstract}
We study the ground state properties of a quasi one dimensional electron gas,
interacting via an effective potential with a harmonic transversal
confinement and long range Coulomb tail. 
The exact correlation energy has been calculated for a wide range of electron
densities by using the lattice regularized diffusion Monte Carlo method, 
which is a recent development of the standard projection Monte Carlo
technique. In this case it is particularly useful 
as it allows to sample the exact 
ground state of the system, even in the low density regime 
when the exchange between electrons is extremely small.
For different values of the width parameter $b$ ($0.1~a^*_0 \le b \le 4
~a^*_0$), we give a simple parametrization of the correlation
energy, which provides an accurate local density energy functional for quasi
one dimensional systems.
Moreover we show that static correlations are in qualitative agreement
with those obtained for the Luttinger liquid model 
with long range interactions.
\end{abstract}

\maketitle

\section{Introduction}
The recent experimental realizations of ideally clean quasi one dimensional
(Q1D) systems, like ultra cold Fermi gases in elongated harmonic traps
\cite{cold_atoms} and high mobility quantum wires in the so-called
cleaved edge overgrowth samples\cite{ceo}, 
have stimulated a new intense theoretical
effort to explain the physical outcome of these systems. The low
dimensionality brings about peculiar phenomena such as the fractionalized (0.7
structure) conductance \cite{fraccond,auslaender05}, 
enhances the effect of localization (Wigner
crystallization)\cite{auslaender05,localization}, 
breaks the validity of the Fermi-liquid paradigm, which must
be abandoned in favor of the Tomonaga-Luttinger liquid (TLL)
concept\cite{lutt}.
 One of the most striking consequences of the Luttinger theory is the
 spin-charge 
separation, which has been seen in a series of remarkable experiments carried
out by Auslaender \emph{et al.}, who were able to resolve the
dispersion energy of elementary spin and charge excitations
\cite{auslaender02,auslaender05}. They used the
tunneling current between two parallel wires to probe the properties within one
of the two wires. However, their measurements exhibit some features, like
fringes in the tunneling pattern and non unitary conductance, that are not
completely understood.

The finite-size effects, the disorder 
and the inhomogeneity of the device can play a crucial
role to quantitatively explain the experiments\cite{matveev,mueller,fiete}. 
In this paper we will rather
focus on the simpler homogeneous Q1D electron gas with harmonic transversal
confinement and effective interactions with a long range Coulomb tail ($1/r$). 
The details of the confinement only affect the behavior at short range of the
effective potential, and many models\cite{harm,hard,2dharm,2dhard,2dcoul} 
have been proposed which give an equivalent description of the homogeneous Q1D
Coulomb gas. 
Despite its simplicity and a huge amount of theoretical
work\cite{schulz,parola,fratini,capponi,exact} done to understand 
its properties, an accurate parametrization of its correlation energy is still
missing. Indeed, the calculation of the ground state energy of 1D wires with
realistic Coulomb interactions is still an open problem, since the TLL
is an effective low energy theory, and the
RPA perturbative expansion is correct only in the high density limit. 
Recently a mapping of the problem with realistic Coulomb interaction onto
exactly solvable models has been proposed\cite{fogler}, 
but within this scheme several approximations are required for different
density regimes. Previously an STLS-like\cite{gold}
method was used by Calmels and Gold\cite{bloch,calmels} 
to compute the correlation energy, but it turns out to be not accurate enough 
to yield the correct ground state in the low density regime.
Indeed, it predicts a Bloch instability ruled out by the Lieb-Mattis
theorem\cite{lieb}.

Projection quantum Monte Carlo (QMC) 
techniques are exact in one dimension, since
in this case the ground state (GS) nodes are known exactly\cite{ceperley91}, 
and the so called fixed node approximation, 
which cures the well known sign problem, 
does not affect the results. However, previous diffusion Monte Carlo (DMC)
simulations\cite{malatesta} suffered from a lack of ergodicity at low electron
densities, when the exchange between electrons become exponentially small. 
Other QMC attempts\cite{macdonald} 
to study the Q1D electron gas used the ``world-line''
algorithm on the lattice after a naive discretization of the
Laplacian. Here we apply the novel lattice regularized diffusion Monte Carlo
(LRDMC)\cite{lrdmc}, which is more efficient than previous projection 
QMC methods at low density, where it substantially alleviates the lack of
exchanges between electrons. 

The aim of this work is to provide a simple and efficient parametrization of
the ground state correlation energy, exactly computed for the unpolarized
system. Other ground state properties are also studied, like the spin and
charge structure factors, which reveal a strong similarity with those computed
using a TLL with long range interactions\cite{schulz}. 
Therefore our parametrization can be an extremely useful input
of density functional theory (DFT) calculations of Q1D systems with local
density approximation (LDA), which can include the homogeneous Q1D electron gas 
with realistic long range interactions and TLL features as the reference
system. Previous successful attempts have been limited so far to model systems 
such as the  Luttinger liquid\cite{dft_luttinger} 
or the one dimensional  Hubbard model\cite{dft_luther},
that have been successfully 
used as the reference systems for DFT simulations of Q1D ultra cold
inhomogeneous atomic gases. 
We believe that an essentially exact calculation of the correlation energy, 
presented in this paper, should open the way for 
a wide range of relevant realistic applications 
in the field of Q1D systems. 
 
The paper is organized as follows. In Chap.~\ref{model} we present the model
for the Q1D homogeneous electron gas, in Chap.~\ref{wave_function} we
describe the variational ansatz used in our simulations, in Chap.~\ref{LRDMC}
we briefly review the LRDMC method and we compare it with the standard DMC
algorithm. The results are reported in Chap.~\ref{correlation_energy}, where
we present the parametrization of the correlation energy with different values
of the transversal confinement, and in Chap.~\ref{pair_correlations}, where we
study 
the charge and spin structure factors. Finally, the conclusions are drawn in
Chap.~\ref{conclusions}. In the end, 
two appendices explain how to compute the RPA
correlation energy at high densities, and how to estimate the plasmon
excitations from the knowledge of the static structure factor. 

\section{Model}
\label{model}
In this paper we study a realistic model for a quantum wire with the 
lateral confinement provided by a harmonic transversal potential $V(r_\bot)
= \frac{r_\bot^2}{4 b^4}$, where $b$ tunes the strength of the confinement
and measures the wire width. Here and henceforth we use the effective Bohr
radius $a_0^*=\frac{\hbar^2 \epsilon}{m^* e^2}$ as length unit and the
effective Rydberg $Ryd^*=\frac{e^2}{2 \epsilon a_0^*}$ as energy unit, 
where $\epsilon$ is the dielectric constant of the
semiconducting medium and $m^*$ is the effective mass of the electrons in the
semiconductor. The electrons in the wire interact via a long range 
Coulomb potential. 
If the confinement is sufficiently strong, the ground state (GS) of
this system can be approximated with good accuracy by a wave function 
with longitudinal and transversal components factorized. In
particular we neglect any contribution from higher subbands of the lateral
direction and we take the GS of the two dimensional harmonic
oscillator as the transversal part of the total wave function. This
approximation is valid whenever 
\begin{equation}
r_s >> \frac{\pi b}{4},
\end{equation}
where $r_s$ is the Wigner-Seitz radius ($2 r_s = 1/\rho$ is the mean interparticle
distance), i.e. for sufficiently low electronic density. 

Tracing out the transverse motion from the full Schr\"odinger equation by
integration over the lateral coordinates of the particles yields\cite{harm},
for $N$ particle on a segment of length $L$, the
one dimensional (1D) Hamiltonian 
\begin{equation}
\label{ham}
H = -\sum_{i=1}^N \nabla^2_i + \frac{1}{2L}\sum_{k\ne 0} \tilde{V}_b(k)
\left[\rho(k)\rho(-k)-N\right] 
\end{equation}
with $\rho(k)=\sum_j \exp(i kx_j)$  the Fourier transform of the one-body density operator and  
\begin{equation}
\tilde{V}_b(k)=2 E_1(b^2 k^2) \exp(b^2 k^2).
\label{Fourier_potential}
\end{equation}
Above 
$E_1$ is the exponential integral function and in Eq. (\ref{ham})
  a suitable rigid  positive charge background exactly cancels the $k=0$ term.
The real space   1D  interparticle potential,  
\begin{equation}
V_b(x)= \frac{\sqrt{\pi}}{b} {\rm exp} \left( \frac{x^2}{4 b^2} \right) {\rm
  erfc} \left(\frac{ |x|}{2 b} \right),   
\end{equation}
has a long-range Coulomb tail but is finite at the origin.

Since in this work we focus on the GS properties of the wire, 
we work  in the sector of vanishing total spin component along the $z$ 
quantization axis, namely 
$N^\uparrow=N^\downarrow=N/2$. Obviously  this choice 
 is not a restriction  because the ground  state certainly belongs 
to this sector, regardless of its total spin. 
On the other hand the value of the total spin is known 
from the Lieb-Mattis theorem and Ref.~\onlinecite{malatesta}: 
the GS of the one dimensional Coulomb gas 
is always a singlet for \emph{all} densities.
In order to perform a finite-size scaling analysis of the 
energy and correlation functions, we carried out QMC
simulations with different number of particles, going from $N=10$ to
$N=162$ and with an odd number of particles per each spin so that
the degeneracy effects are avoided. With the aim to
further reduce the finite-size bias, we considered the Hamiltonian in
Eq.~\ref{ham} with periodic boundary conditions (PBC) and with an infinite
number of replicas of the simulation box (supercell). Thus, an electron in a
supercell interacts with the other electrons in the supercell, their
images,  its own images, and the background. It is then convenient  to define  
an effective interparticle potential, by summing  the bare interaction  of a
particle and its background with a second particle over all its images, 
to obtain a periodic function:      
\begin{eqnarray}
\label{sum}
V(x)&=&\sum_n \left[V_b(x+n L)-\frac{1}{L}\int_{-L/2}^{L/2}dy~V_b(x+nL-y)\right]
\nonumber \\
&=&\frac{1}{L}\sum_{n\ne 0}\tilde{V}_b(G_n)e^{\imath G_n x}.
\end{eqnarray}
In the above expression 
$L$ is the length of the simulation box, $n$ takes relative
integer values and $G_n=2\pi n/L$ is a reciprocal vector
of the 1D Bravais lattice with primitive unit cell of length $L$.
Since $V_b$ is a
long-range potential, we resort to an  Ewald-like method\cite{allen} to
compute the sum in Eq.~\ref{sum}.  The short-range part of the
potential $V$ and its long-range tail are treated in a different
fashion, the former being summed in the direct space, the latter in
the reciprocal one.  The Ewald's procedure yields:
\begin{widetext}
\begin{eqnarray}
V(x)& = & V_{\rm sr}(x) + V_{\rm lr}(x), 
 \\
V_{\rm sr}(x) & = & \frac{\sqrt{\pi }}{b} 
\sum_{n=-\infty}^{+\infty} \exp \left[ \left(x-n L\right)^2/4b^2\right] 
{\rm erfc} \left( \frac{\left| x-n L\right| }{2b}\right) 
- \sum_{n=-\infty}^{+\infty} \frac 2{\left|x-n L\right| }
{\rm erf}\left( \frac{\left|x-n L\right|}{2b}\right), 
 \\
V_{\rm lr}(x) & = & 2 \sum_{n>0} \frac{\cos (G_nx)}{L} e^{-(bG_n)^2}
\tilde{V}_b(G_n).  
\end{eqnarray}
\end{widetext} 

In practice, we have worked with the Hamiltonian
\begin{equation}
\label{ham1}
H = -\sum_{i=1}^N \nabla^2_i + \sum_{i<j} V(x_{ij})+\frac{N}{2}V_{\rm MAD},
\end{equation}
where $V_{\rm MAD}=V(0)-V_b(0)$ is the Madelung energy, i.e., the
interaction of a particle with its own images.  We have used a
tabulation for the potential $V(x)$. In particular, the $G$ sum in
$V_{\rm lr}$ has been truncated at $G=12/b$ and a sufficient number of
images in $V_{\rm sr}$ has been included, so that the overall error in the
tabulation is less than $10^{-6} Ryd^*$.

\section{Wave function}
\label{wave_function}
The wave function $\Psi_T$ we used in our QMC simulations is of 
the Slater-Jastrow type:
\begin{equation}
\label{wf}
\Psi_T= \exp \left(-\sum_{i<j} u(x_{ij}) \right) D^\uparrow D^\downarrow,
\end{equation}
where $D^\sigma$ is a determinant of $N^\sigma$ plane waves with wave vectors
occupied up to the Fermi momentum $k_F=\frac{\pi}{4 r_s}$. 
The Jastrow factor contributes significantly to improve 
the quality of the variational state, since it correlates the particles and
tunes the amplitude of $\Psi_T$. 
Here we use a two body Jastrow factor, which takes into account the
electron-electron correlation, without spoiling the translational invariance
of the system. A recent work \cite{capello} 
on the 1D $t-t^\prime$ Hubbard model has shown
that the long-range behavior of the two body Jastrow effectively accounts for
the proper description of the metallic and insulating phases of that lattice
model. Therefore a good functional form of $u(x)$ in Eq.~\ref{wf} 
is crucial to obtain the correct physics for a strongly correlated 1D system.

In order to avoid spin contamination\cite{umrigar}, the 
function $u(x)$ does not depend on the spin of the particles.
In particular, we choose the RPA form of $u(x)$ as our first 
variational ansatz.
Following Ref.~\onlinecite{gaskell}, the Fourier components of $u$ are: 
\begin{equation}
2 \rho u_{RPA}(k)= -S_0(k)^{-1}+\sqrt{S_0(k)^{-2}+ 2 \rho V_b(k)/k^2}, 
\label{u_rpa}
\end{equation}
with $S_0(k)=(k/2k_F)\theta(2k_F-k)+\theta(k-2k_F)$ the structure factor of a
non interacting one dimensional electron gas (1DEG).
The two body Jastrow is repulsive at the origin, to reduce the
probability of two electrons to approach each other, 
and lower in this way the potential energy.
Since the effective Coulomb coupling increases as the density decreases,
the lower is the density, the higher  is the Jastrow repulsion.

Rescaling $u_{RPA}$ is the most straightforward improvement beyond the RPA
ansatz. It improves the variational wave function by
means of the simple parametrization
\begin{equation}
u(x)=\gamma u_{RPA}(x),
\label{u_rpa_opt}
\end{equation}
where $\gamma$ is a linear parameter that we optimized using the variance
minimization\cite{firstvariance,cyrusvariance}. 
The optimal values for $\gamma$ are reported in Tab.~\ref{scaled} for different densities 
and for $N=22$: the RPA seems good for the highest densities,
where $\gamma \simeq 1$, but it becomes worse for lower densities, 
where the rescaling is effective. Indeed, whenever the correlation is stronger, 
$u_{RPA}$ overestimates the interelectron repulsion. 
Moreover we have found that the optimal value of $\gamma$ does not
depend on the number of particles in the supercell,
the function $u_{RPA}$ having the proper dependence on $N$. 
\begin{table}[!htp]
\caption{\label{scaled}
Scaling parameter $\gamma$ of the $u_{RPA}$ function optimized using the
method of variance minimization for the 1DEG with $b=0.1$ and $N=22$}
\begin{ruledtabular}
\begin{tabular}{l|l|l|l|l|l }
& \multicolumn{1}{l}{$r_s=1$}
& \multicolumn{1}{l}{$r_s=2$}
& \multicolumn{1}{l}{$r_s=4$}
& \multicolumn{1}{l}{$r_s=6$}
& \multicolumn{1}{l}{$r_s=10$}
 \\
\hline
$\gamma$   & 1.05 & 1.00 & 0.83 & 0.67 & 0.61  \\
\end{tabular} 
\end{ruledtabular}
\end{table}

\begin{figure}[!ht]
\centering
\includegraphics[width=\columnwidth]{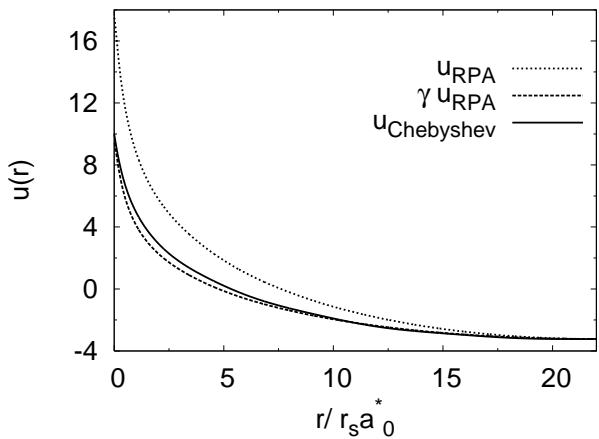}
\caption{Optimized $u$ functions for $r_s=10$ and $N=22$:
$u_{RPA}(x)$ (dotted line), $\gamma u_{RPA}(x)$ (dashed line), and the
Chebyshev expansion for $u(x)$ (solid line). $r_s a_0^*$ is the unit
length.}   
\label{comparison}
\end{figure}

With the aim to check 
whether the scaled RPA is accurate enough to yield the correct Jastrow
correlation for the lowest densities, when the effective coupling is higher, 
we take into account the most general expression for the 
two body Jastrow. We expand $u$ in a linear sum of Chebyshev
polynomials\cite{williamson}, which
are a complete basis set in the orthogonality interval $(-1,1)$:
\begin{equation}
u(x) = \sum_{m > 0} \alpha_m T_{2m}
 \left( \frac{x-\frac{L}{2}}{\frac{L}{2}} \right),
\label{tchebymin}
\end{equation}
where the range $(0,\frac{L}{2})$ is mapped into the interval $(-1,0)$ and
only the even polynomials are used. In this way the condition
$u^\prime(L/2)=0$ is fulfilled, and ensures the continuity of the first
derivative of $u$ at the edge of the supercell. 
Moreover the sum in Eq.~\ref{tchebymin} starts
from $m=1$, since $T_0$ is the identity and the wave function is
determined apart from a constant.
Using the variance minimization, 
we optimize the parameters $\alpha_m$ up to the convergence of the expansion
in Eq.~\ref{tchebymin}, which has been reached for $m=10$. 
The variational energies relative to the
various functional forms of $u$ for $r_s=10$ and $N=22$ are summarized in
Tab.~\ref{compenergy}, whereas the functions are plotted in  
Fig.~\ref{comparison}. 
The comparison shows that the scaled RPA Jastrow factor
leads to a very good variational state, 
as its energy is very close to the exact GS
value (more than $99.5 \%$ of correlation energy is recovered) and it is almost
coincident to  the most flexible variational form obtained with  
the mentioned  Chebyshev expansion. 
The simple RPA form is a good approximation, since it
provides a large fraction of the correlation energy ($98.3 \%$), but the 
rescaling of the RPA Jastrow yields a further substantial improvement. Thus
it is clear that
the most convenient parametrization of the Jastrow term is the scaled
$u_{RPA}$, because one is able to reach an almost exact wave
function already at the variational level, by optimizing just one variational
parameter. 
For the above reason, in the forthcoming sections we will use the
optimized and rescaled RPA Jastrow as trial wave function $\Psi_T$
for all densities ($r_s$) and sizes ($N$) taken into account in  
this QMC study of the 1DEG.

\begin{table}[t]
\caption{Total energy $E_{tot}$, correlation energy $E_{corr}$ and percentage of
correlation energy $\% E_{corr}$ 
for $b=0.1$, $r_s=10$, and $N=22$. The fraction of
the correlation energy recovered is computed from LRDMC
calculations which provide the exact GS energy for a 1DEG (see
Sec.~\ref{LRDMC}).} 
\label{compenergy}  
\begin{ruledtabular}
\begin{tabular}{l|l|l|l}
         & \multicolumn{1}{l}{$E_{tot}$}
         & \multicolumn{1}{l}{$E_{corr}$}
         & \multicolumn{1}{l}{$\% E_{corr}$}  \\
\hline
RPA                         & -0.47207(2)  & -0.20519(56)   &  0.9830(15) \\
\hline
Scaled RPA                  & -0.474825(9) & -0.20794(55)   &  0.9962(15) \\
\hline
Chebyschev                 & -0.474900(9) & -0.20802(55)   &   0.9965(15) \\
\end{tabular}
\end{ruledtabular}
\end{table}

\section{Lattice regularized Diffusion Monte Carlo method}
\label{LRDMC}

The low dimensionality and the strong correlation among the electrons not only
 have dramatic consequences on the physical properties of the quantum wire,
which will be studied in Secs.~\ref{correlation_energy} and
 \ref{pair_correlations}, 
but also affect the efficiency of the QMC simulations of the system.
Indeed, as we have seen in the previous
section, the effective Coulomb interaction leads to an optimal wave function
with a strong repulsive Jastrow factor, which freezes the relative positions of the
particles and enhances  the $2 k_F$ component of the
charge-charge correlation, a signature that the system is close to the
Wigner phase.
The two body Jastrow, necessary to provide a good variational
description of the system, introduces \emph{pseudo nodes}, i.e. surfaces in the
configuration space where the wave function almost vanishes, due to the
exponentially increase of $J(x)$, which acts like a Gutzwiller projector, by
avoiding ``double'' occupancies on a given electronic position $x$. 
These \emph{pseudo nodes} are similar to the usual nodes of the fermionic wave
function, but if the latter arise from the antisymmetrization of the many body
state, the former are a consequence of the strong repulsion, which prevents two
electrons to come closer and eventually overlap. The effect of the
\emph{pseudo nodes} on the QMC simulation is harmful, since 
in the 1D case they can lead to a slow convergence of the Markov chain to the
equilibrium distribution.
In particular, it can be extremely difficult to connect two configurations with
a spin exchange. However, the charge degrees of freedom are still well
reproduced in spite of this lack of ergodicity, as one might infer that 
the pseudo nodal pockets are
equivalent for the charge properties. Instead the expectation values of 
spin dependent operators are spoiled, if the Markov
chain is not able to guarantee a sufficient number of spin exchanges during
the simulation in a reasonable time.

The variational Monte Carlo (VMC) algorithm can easily overcome the problem,
since the proposed move can be forced to flip the spin of an electronic
configuration, either 
by explicitly introducing a spin exchange or by allowing the
amplitude of the move to be greater than the mean interparticle
distance. Instead in 
the diffusion Monte Carlo (DMC) approach, the random walk has to follow the
diffusion process driven by the imaginary time dependent Schr\"odinger
equation. If the importance sampling is introduced,
the resulting Green function, approximated by means of the Trotter
expansion up to the first order in the time step $\tau$, 
includes the drift-diffusion dynamics:
\begin{equation}
{\bf R}^\prime= {\bf R} + D \tau \nabla \ln |\Psi_T({\bf R})|^2 + \chi \sqrt{2 D
  \tau},
\label{displacement}
\end{equation}
where $\nabla \ln |\Psi_T({\bf R})|^2$ is the quantum force, $D=1$ is
the diffusion coefficient, and $\chi$ is a Gaussian distributed random
variable. The configurations
generated step by step are distributed accordingly to $|\Psi_T|^2$
at the beginning of the simulation, but after a transient they will 
reach the equilibrium and sample the 
mixed distribution $\Psi_{FN} \Psi_T$, where $\Psi_{FN}$ is the 
lowest variational state with the same nodes as the trial wave function
$\Psi_T$ (this is the so called fixed node (FN) constraint).
In order to get rid of the time step bias in the final result, one
needs to extrapolate the FN energies obtained at different time steps
for $\tau$ going to zero. For smaller $\tau$, spin exchanges become rarer,
because the mean square displacement vanishes linearly with $\tau$
(Eq.~\ref{displacement}). 
To overcome the lack of spin exchanges in the DMC, we have applied
a different projection QMC method, the lattice regularized Green function Monte
Carlo (LRDMC), successfully introduced 
in Ref.~\onlinecite{lrdmc} to cure the localization error in the presence of
non local potentials. In this section we review the method and 
compare its efficiency of sampling spin flips with respect to the standard DMC
framework.

The main idea behind the LRDMC 
method is to deal with a regularized Hamiltonian in
such a way that the standard
Green function Monte Carlo (GFMC) algorithm\cite{ceperley,calandra,capriotti} 
on a lattice can be applied also to continuous systems. 
The regularization of the Hamiltonian in Eq.~\ref{ham} involves both the
kinetic and the potential parts.
The Laplacian is discretized by means of the finite
differences
\begin{equation}
\Delta = \eta \left [ p \Delta^a +
  (1-p) \Delta^{a^\prime} \right ] +O(a^2),
\label{kinetic}
\end{equation} 
with $\Delta^a$ an Hermitian lattice operator given by 
\begin{equation} \label{laplace}
\Delta^a \Psi(x_i) = \frac{1}{a^2} 
\left( \Psi(x_i+a) + \Psi(x_i-a) - 2 \Psi(x_i) \right ),
\end{equation}
where $a$ is the mesh size, $p$ and $\eta$ are constants ($\eta=1+O(a^2)$),
and $x_i$ is the position of the $i$-th electron. 
Due to the homogeneity of
the system, $p$ is kept spatially independent, contrary to the general
case\cite{lrdmc}  where the dependence of $p$ on the electronic positions 
is exploited to improve the efficiency of the diffusion process. 
Here $p=0.5$, and the contributions to the total Laplacian coming from
$\Delta^a$ and $\Delta^{a^\prime}$ are equally weighted. 
The two terms, with $a^\prime/a=\sqrt{5}$, 
allow the diffusion to explore all the continuous space, since the two meshes
are incommensurate; in this way the lattice space bias due to the
discretization of the continuous kinetic operator 
is greatly reduced and one can work
with a reasonably large value of $a$ without a significant lattice step error.

Also the potential is regularized, so that
our final Hamiltonian $H^a$ fulfills the following three conditions: 
i) $H^a \to H$ for $a \to 0$;
ii) for the  chosen  trial wave function $\Psi_T$,
for any $a$ and any configuration $x$,
the local energy $e_L ({\bf R},[\Psi_T])=
{  H \Psi_T ( {\bf R} \over \Psi_T  ({\bf R}}$ of the continuous 
Hamiltonian $H$ is equal to the one 
 $e^a_L({\bf R},{[\Psi_T]})$ corresponding 
to the Hamiltonian $H^a$;
iii) the discretized kinetic energy is equal to 
the continuous one calculated on the state  $\Psi_T$.
The condition (iii) determines the constant $\eta$, 
while the condition (ii) fixes the form of the regularized 
potential $V^a$: 
\begin{equation} \label{secondreg}
 V^a({\bf R}) = V({\bf R})+  \frac{1}{2 } \left[ \frac{\sum_i ( \Delta^a_i -
     \Delta_i) \Psi_T} {\Psi_T}\right]({\bf R}).
\end{equation}
Notice that the condition (ii) yields another important property
for $H^a$: if $\Psi_T$ is an eigenstate of $H$, it is also  
an eigenstate of $H^a$ for any $a$. 
Thus, as the quality of $\Psi_T$ increases, the dependence of the LRDMC 
energy on $a$ decreases.

The lattice regularized Hamiltonian $H^a$ reads:
\begin{equation}
H^a_{{\bf R'},{\bf R}} = \left \{
\begin{array}{ll}
-\eta ~ p/a^2 & \textrm{if ${\bf R'} = {\bf R} + \delta_a$} \\
-\eta ~ (1-p)/{a^\prime}^2 & \textrm{if ${\bf R'} = {\bf R} + \delta_{a^\prime}$} \\
2N \eta \left( \frac{p}{a^2} + \frac{1 - p}{{a^\prime}^2} \right) + V^a({\bf R}) &
 \textrm{if ${\bf R'} = {\bf R}$},
\end{array} 
\right.
\label{h_discretized}
\end{equation}
where $\delta_a$ ($\delta_{a^\prime}$) 
is a $N$ dimensional vector defined as the \emph{one} particle
displacement of length $\pm a$ ($\pm a^\prime$). Thus
there are $2N$ different $\delta_a$ ($\delta_{a^\prime}$),
and $H^a$ in Eq.~\ref{h_discretized} contains $4N$ off diagonal elements,
which come from the discretization of the Laplacian (Eq.~\ref{kinetic}).
In particular, by defining the importance sampling Green function,  
$G_{{\bf R'},{\bf R}} = \Psi_T({\bf R'}) (\Lambda \delta_{{\bf R'},{\bf R}}-
H_{{\bf R'},{\bf R}})/\Psi_T({\bf R})$, the configuration ${\bf R}$ is
connected by $G_{{\bf R'},{\bf R}}$ 
to a \emph{finite} number of configurations ${\bf R'}$, although
${\bf R}$ and ${\bf R'}$ live in a continuous space. 
Therefore the Green function $G_{{\bf R'},{\bf R}}$ is \emph{discrete},
and can be sampled using a heat bath algorithm, like in the standard GFMC
scheme on a lattice, although in this case ${\bf R}$ and ${\bf R'}$ are
continuous variables. Another important difference
with respect to the lattice case is the spectrum of $H^a$, which is unbounded
from above; thus, in order to guarantee the positivity of the Green function,
we need to perform the limit $\Lambda\to\infty$, which can be handled within the
continuous time formulation, already introduced in Ref.~\onlinecite{capriotti}
for the GFMC method.

Although in the continuous limit $a\to 0$ there is no sign problem because 
the sampling is restricted within a region -the nodal pocket- with definite 
sign, for non zero $a$ the fermionic sign problem is still
present and needs to be treated by means of an effective Hamiltonian 
\cite{ceperley}, which approximates the regularized $H^a$ (FN constraint). 
The regularized effective Hamiltonian
$H^{\textrm{eff}}$ included in the LRDMC algorithm is defined as follows:
\begin{equation}
H^{\textrm{eff}}_{{\bf R},{\bf R'}} = \left \{
\begin{array}{ll}
H^a_{{\bf R},{\bf R'}}                & \textrm{if ${\bf R} \ne {\bf R'}$ and} \\
 & \textrm{$\Psi_T({\bf R'}) H^a_{{\bf R},{\bf R'}} / \Psi_T({\bf R}) \le 0$} \\
0                             & \textrm{if ${\bf R} \ne {\bf R'}$ and} \\
 & \textrm{$\Psi_T({\bf R'}) H^a_{{\bf R},{\bf R'}} / \Psi_T({\bf R}) > 0$} \\
H^a_{{\bf R},{\bf R}} + \mathcal{V}_{sf}({\bf R}) & \textrm{if ${\bf R} = {\bf R'}$}, \\
\end{array} 
\right. 
\label{effective_fn_hamiltonian}
\end{equation}
where ${\cal V}_{sf} ({\bf R})=\sum_{{\bf R'} \ne {\bf R} } \Psi_T({\bf R'})
H_{{\bf R'},{\bf R}}^a/\Psi_T ({\bf R}) > 0$, the so called 
sign-flip term, is the sum over all the terms that cause  
a negative sign  problem in the Monte Carlo  sampling.
For this reason these 
terms are traced in the diagonal part of the effective  Hamiltonian, 
that therefore no longer contains off diagonal terms with the
''wrong'' sign.
The ground state $\Psi_{FN}$ of $H^{\textrm{eff}}$ 
has the same signs as the trial wave function $\Psi_T$, and so the
mixed distribution $\Psi_{FN} \Psi_T$ sampled during the LRDMC simulation 
will be non negative. In general, in the limit $a \rightarrow 0$ the FN energy
$E_{FN}=\langle \Psi_{FN} | H^{\textrm{eff}} | \Psi_T \rangle$ of the
effective Hamiltonian $H^{\textrm{eff}}$ is an upper bound of the GS
energy $E_0$ of $H$. 
The FN approximation turns out to be exact 
only if the nodes of the trial wave function are the same as the GS nodes. 
However, the trial wave function $\Psi_T$ in Eq.~\ref{wf} has the exact
GS nodes. Indeed in the 1D case the nodal structure of the GS
is exactly defined by the coalescence planes $x_i=x_j$, where $x_i$ and
$x_j$ are two electrons with the same spin,
and the position of the planes are completely determined by the antisymmetry of
the particles \cite{ceperley91}.
Therefore both the LRDMC ($a \rightarrow 0$) and DMC 
($\tau \rightarrow 0$) values for this 1D system are exact within their
statistical precision. 

We did an accurate comparison between the DMC and LRDMC approaches, by taking
into account the efficiency of the energy estimate, the dependence on the
time step and on the lattice space, and the spin flip frequency 
during the simulations, defined as the number of exchanges between two
particles with opposite spin per unit time (imaginary projection time)
per particle.  
The ``standard'' DMC algorithm, taken as the reference
for our comparison, is described in Ref.~\onlinecite{umrigar93}.
We applied the two QMC schemes to 
the quantum wire model with $N=22$ and for $r_s$ ranging from $1$ to $10$. For
a fair comparison, we chose the DMC time step $\tau=a^2/2 $, so that 
both algorithms provide the same amplitude for the diffusion move.
The efficiency of the DMC energy estimate is
double that of the LRDMC, since in the latter approach
we need to compute in advance all the possible off diagonal moves, thus losing a
fraction of the computing time. On the other hand, as reported in Table
~\ref{exchange}, the spin exchange frequency
is almost the same for the high density model, when the correlation is
weak, but the LRDMC becomes more and more effective in sampling the spin flips
when the density is lower and the correlation becomes stronger. In
particular, for the lowest density ($r_s=10$) 
the LRDMC algorithm yields an efficiency in
the spin flip sampling which is 
two orders of magnitude higher than the DMC case.

\begin{table*}
\caption{\label{exchange}
Spin exchange frequency ($Ryd^*$) for the LRDMC and DMC algorithm at
different densities for the quantum wire model with $N=22$ and $b=0.1$.
Notice that the frequency is reduced when the density decreases, while the
efficiency of the LRDMC increases with respect to the DMC. All the simulations
have been performed with $a=0.2 r_s$ and $\tau=a^2/~2$. }
\begin{ruledtabular}
\begin{tabular}{l|d|d|d|d|d }
& \multicolumn{1}{c}{$r_s=1$}
& \multicolumn{1}{c}{$r_s=2$}
& \multicolumn{1}{c}{$r_s=4$}
& \multicolumn{1}{c}{$r_s=6$}
& \multicolumn{1}{c}{$r_s=10$}
 \\
\hline
LRDMC  &  1.18 & 5.72~\mbox{$10^{-2}$} & 1.96~\mbox{$10^{-2}$} &
3.19~\mbox{$10^{-4}$} &  8.86~\mbox{$10^{-6}$} \\
\hline
DMC  &  1.14 &  3.81~\mbox{$10^{-2}$} & 4.24~\mbox{$10^{-3}$} & 
3.21~\mbox{$10^{-5}$}  & 9.09~\mbox{$10^{-8}$} \\
\hline
rel eff &  1.03 & 1.50 & 4.62 & 9.94 & 97.47 \\
\end{tabular} 
\end{ruledtabular}
\end{table*}

\begin{figure}[!ht]
\centering
\includegraphics[width=\columnwidth]{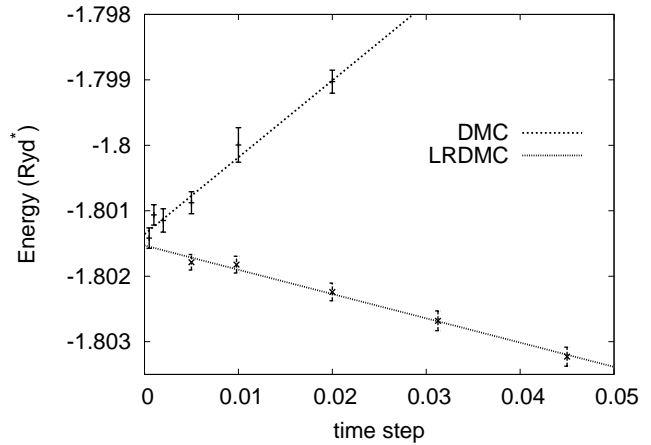}
\caption{
DMC and LRDMC energies dependence on the time step ($b=0.1$,$r_s=1$,$N=22$). 
The lattice space $a$ has
been mapped into the time step, by means of the relation $a=\sqrt{2 \tau}$.
For both the two cases, the dependence appears to be linear, 
with a slope of 0.117(8) for the DMC algorithm and -0.018(2) for the LRDMC approach.}
\label{error_qw}
\end{figure}

Another appealing behavior of the LRDMC approach is the lattice space
dependence of the fixed node energy. As one can see in Fig.~\ref{error_qw}, the
LRDMC energies have a quadratic dependence on $a$  
with a prefactor much smaller than the slope of the linear fit
for the corresponding DMC energies. This means that in order to obtain an almost
converged LRDMC result one does not need to go to small lattice spaces, with a
gain both from the computational point of view and from the efficiency in  the
spin flip sampling, 
which of course is reduced as the diffusion move goes to zero.

To summarize, it is apparent that both the lattice step bias and the lack of
ergodicity are greatly reduced by using the LRDMC algorithm in the place of the
standard DMC. Therefore, given the amplitude of the QMC move, 
the LRDMC is more effective than the DMC scheme. 
We believe that the reason is related to the Trotter
approximation behind the DMC propagator, which spoils the exact dynamics of
the diffusion process and apparently affects the ergodicity of the random
walk. On the other hand, even within  a finite  lattice space, 
the LRDMC algorithm converges  to the exact GS of the  
effective Hamiltonian $H^{\textrm{eff}}$ in
Eq.~\ref{effective_fn_hamiltonian},  implying that simulations 
are always physically meaningful even for  $a>0$.
This should  allow in general  a more 
controlled and smooth 
extrapolation to  the continuous $a\to 0$ limit.

In the next section, we will analyze both the VMC and the LRDMC
results for the quantum wire model. For every density and width, we have
performed a lattice space extrapolation ($a \to 0$) and a finite size
extrapolation to the thermodynamic limit ($N \to \infty$). To extrapolate the
energies to the continuous limit, we have fitted points computed 
in the range $0.05 \le a \le 1.2$ with a quadratic function in $a$, as reported in
Fig.~\ref{error_qw}, while for the extrapolation to the thermodynamic limit we
used the function $E_\infty + c_1/N + c_2/N^2$ in our fits. 
We evaluated the LRDMC energies 
for $N=10,22,42,62,82$, but in some cases, for low $r_s$, we carried out
LRDMC simulations with as many as $242$ particles, in order to have always 
a reliable estimate of the finite size errors.

\section{Correlation energy}
\label{correlation_energy}

The correlation energy per particle $E_{corr}=E_0-E_{HF}$ 
 is computed
for various values of the width parameter $b$ and densities $r_s$, 
and parametrized as a function of $r_s$ for each $b$.
The unpolarized HF energy of this quantum wire model is
\begin{equation}
\label{HF_energy}
E_{HF}(r_s,b)=\frac{\pi}{48 r_s^2} +  F(\frac{\alpha r_s}{2 b})/b,
\end{equation}
where the first term is the kinetic energy, and the second contribution 
is the exchange energy ($E_{ex}$) term with $\alpha=4/\pi$ and the function
$F$ defined by 
\begin{equation}
F(R) = - \frac{1}{2 \pi} \int_0^{1/R} dx f(x) \left [ 1 - R x  \right ].
\end{equation}
The Hartree term vanishes since the system is neutral and homogeneous.

We evaluated the GS energy $E_0$ using the LRDMC method with the FN
approximation, which projects the initial $\Psi_T$ to the lowest
energy state of the system with the same nodes of $\Psi_T$.
However, since the trial wave function
in Eq.~\ref{wf} has the exact GS nodes, our LRDMC energies are exact within
their statistical precision in the limit $a \to 0$, 
as already pointed out in Sec.~\ref{LRDMC}. 
The correlation energy is then obtained by subtracting the $HF$ energy of
Eq.~\ref{HF_energy} from the GS energy. 
Thus, the correlation energy is computed exactly for a given value
of $r_s$ and $b$.

\begin{figure}[!ht]
\centering
\includegraphics[width=\columnwidth]{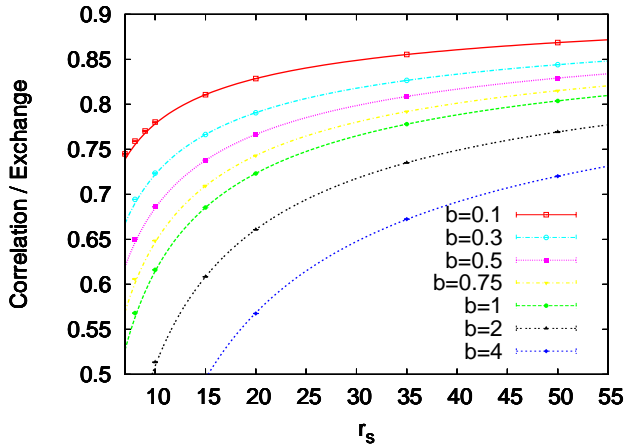}
\caption{(Color online) Values of $E_{corr}/E_{ex}$ plotted versus $r_s$ 
in the low density regime for various thickness $b$. 
The curves are obtained from a fit of the LRDMC
points using the function in Eq.~\ref{fit_ratio} in the range $15 \le r_s \le 50$.}
\label{exchange_corr}
\end{figure}

It is useful to study the correlation energy 
in the high and low density limits, in order to find a good parametrization
which includes the correct asymptotic behavior.
In the high density limit, i.e. $r_s \rightarrow 0$, the correlation energy
can be computed via a perturbative expansion of the interaction, using the RPA
technique to find the coefficient of the lowest order term in $r_s$ (see
Appendix \ref{appendix_RPA}). It turns out that the correlation energy is
quadratically vanishing, as $r_s$ goes to zero
\begin{equation}
\label{high_density}
E_{corr}(r_s \rightarrow 0) = - \frac{A_{corr}}{\pi^4 b^2} r_s^2,
\end{equation}
where $A_{corr}=\int_0^\infty dx~~x f(x)^2  = 4.9348$. This result was 
obtained by Calmels and Gold, using the mean spherical approximation
(MSA)\cite{calmels,gold}, which is consistent with the RPA finding. 
On the other hand, in the low density regime ($r_s \rightarrow \infty$)
the exact behavior of $E_{corr}$ can be guessed by studying the ratio
$E_{corr}/E_{ex}$. For instance, in Ref.~\onlinecite{calmels} the correlation was
computed using a three-sum-rule approach (3SRA) of the STLS theory
for the same model, and this ratio turned out to be
\begin{equation}
\frac{E_{corr}(r_s \rightarrow \infty)}{E_{ex}(r_s \rightarrow \infty)} = 0.84.
\label{en_fraction}
\end{equation}  
Since the asymptotic expansion of $E_{ex}(r_s)$ in the limit of $r_s \rightarrow \infty$ is\cite{calmels} 
\begin{equation}
E_{ex}(r_s) \simeq - \frac{\ln(r_s)}{r_s} \left( \frac{1}{2} + O(\frac{1}{\ln(r_s)}) \right),
\label{asympt_corr}
\end{equation}
it is possible to obtain an asymptotic estimate also for $E_{corr}(r_s \rightarrow \infty)$. 
One can perform the same comparison using the LRDMC data.
In Fig.\ref{exchange_corr} the value of $E_{corr}/E_{ex}$ is
reported as function of $r_s$ for all $b$ studied ($b=0.1,0.3,0.5,0.75,1,2,4$). 
The points have been fitted using the following function:
\begin{equation}
\label{fit_ratio}
\frac{E_{corr}(r_s)}{E_{ex}(r_s)}= \tilde{a} + \frac{\tilde{b}}{\ln(r_s)} +\frac{\tilde{c}}{r_s},
\end{equation}
where $\tilde{a}$, $\tilde{b}$, and $\tilde{c}$ are fitting
parameters. According to Eq.~\ref{asympt_corr}, the
slowest decaying contribution to $E_{corr}/E_{ex}$ should be 
$\propto 1/\ln(r_s)$, which has been included in the fit.
Therefore an accurate extrapolation to $r_s \rightarrow \infty$ is very difficult,
since the logarithmic correction in the low density regime is extremely slow,
and in order to have a reliable estimate of $ \tilde{a}$, one should compute
the correlation energy at much higher values of $r_s$. For instance, in
the 3SRA-STLS theory\cite{calmels}, the ratio
$E_{corr}/E_{ex}$ converges to the value in Eq.~\ref{en_fraction} 
for $r_s > 1000$, which is of course a regime where the QMC framework is not
ergodic. In any case, in the range of densities from $r_s=15$ to $r_s=50$ (the lowest density taken
into account) the function in Eq.~\ref{fit_ratio} fits well our data points, 
apart from the case with $b=4$, which requires even lower densities to enter
into its asymptotic regime. In our fits, $\tilde{a}$ assumes values ranging from $1.0$
to $1.1$, which are slightly larger than those yielded by the STLS theory. 
Thus also our data seem
to support the idea that the correlation energy in quasi-one dimensional
systems behaves as
\begin{equation}
E_{corr}(r_s \rightarrow \infty) \propto -\frac{\ln(r_s)}{r_s},
\end{equation} 
or at least this behavior is compatible with our data until $r_s=50$.

A good parametric representation of the correlation energy has to
satisfy the asymptotic behavior both at $r_s=0$ and $r_s=\infty$. Agosti
\emph{et al.} \cite{agosti} used the interpolation formula
\begin{equation}
\epsilon(r_s)= \frac{r_s^p}{C+D r_s^q},
\end{equation}
with $p$, $q$, $C$, and $D$ variational parameters, in order to fit their STLS
data for a similar 1DEG model system (with hard confinement of a 2DEG in one
direction). They found the above function fits correlation energies very well in
the intermediate range of densities, but it is apparent that their expression
does not yield the ``exact'' asymptotic behavior at high and low densities. 
Here we give a simple analytic representation of the correlation
energy which fits our simulation data over a wide range of densities (from
$r_s=0.05$ to $r_s=50$) and for different values of lateral confinement. Moreover
it includes the correct low and high density limits. 
Following the same lines as Perdew-Wang\cite{perdew}  and Attaccalite \emph{et al.}
\cite{attaccalite}, we use a parametric function which reads 
\begin{equation}
\epsilon(r_s)=-\frac{r_s}{A+ B r_s^n + C r_s^2} \ln ( 1 + \alpha r_s + \beta
r_s^m ).
\label{fit_function}
\end{equation}
The total number of parameters is 7: 3 linear
coefficients ($A$, $B$, and $C$) and one exponent ($0<n<2$) for the polynomial part,
2 linear coefficients ($\alpha$, and $\beta$) and one exponent ($m>1$) in the
argument of the logarithm.
The high density limit of the parametrization is
\begin{equation}
\label{fit_high_density}
\epsilon(r_s \rightarrow 0) = -\frac{\alpha}{A}~ r_s^2,
\end{equation}
and its low density limit is
\begin{equation}
\label{fit_low_density}
\epsilon(r_s \rightarrow \infty) = -\frac{m}{C}~ \frac{\ln(r_s)}{r_s}.
\end{equation}
Thus the expected behavior of the correlation energy is
correctly reproduced by our parametric form (Eq.~\ref{fit_function})
for both the high and low density limits. 
In order to obtain a priori the exact high density result
(Eq.~\ref{high_density}) known from the RPA theory (see Appendix
\ref{appendix_RPA}), we fix the ratio $\alpha/A$
to be equal to $A_{corr}/(\pi^4 b^2)$. Therefore the number of independent
parameters in the function~(\ref{fit_function}) is reduced to 6. In particular,
we determine $A$ from the high density limit, while the other parameters 
are free to minimize the $\chi^2$. Their optimal values are listed in Tab.~\ref{fit_par}. 

\begin{table*}
{\scriptsize
\caption{\label{fit_par} 
  Optimal fit parameters for the correlation energy, as
  parametrized in Eq.~\ref{fit_function}. Different values of $b$ are taken into
  account, for each of them we give its parametrization. In the last rows, we
  report the reduced $\chi^2$ and the overall numerical accuracy $\eta$ in
  $Ryd^*$, defined in Eq.~\ref{fit_accuracy}. $A$ has been determined from the
  high density limit~\ref{high_density}, which is therefore exactly fulfilled
  by our fit. The error of the parameters in the last digits is reported in parenthesis.
}
\begin{ruledtabular}
\begin{tabular}{l d d d d d d d }
 & \multicolumn{1}{c}{$b=0.1$} 
 & \multicolumn{1}{c}{$b=0.3$} 
 & \multicolumn{1}{c}{$b=0.5$} 
 & \multicolumn{1}{c}{$b=0.75$} 
 & \multicolumn{1}{c}{$b=1.0$} 
 & \multicolumn{1}{c}{$b=2.0$} 
 & \multicolumn{1}{c}{$b=4.0$} \\
\hline
$A$  &  4.66(5) &   9.50(9)  & 16.40(11) & 22.53(24)& 32.1(1.3) &  110.5(2.8) &   413.0(5.9) \\
$B$  &  2.092(24) &   1.85(17) & 2.90(13)  & 2.09(12)  & 3.77(23) & 7.90(49)  &  10.8(7)\\
$C$ &  3.735(34) &  5.64(8) & 6.235(46) & 7.363(44) &  7.576(36) &  8.37(6) &  7.99(20) \\
$n$ & 1.379(10) &  0.882(13) &  0.908(7) & 0.906(8) & 0.941(49) &  1.287(26) & 1.549(35) \\
$\alpha$ & 23.63(26) & 5.346(53) &  3.323(22) & 2.029(22) & 1.63(7) &  1.399(35) & 1.308(19) \\
$\beta$ &  109.9(5.5) & 6.69(43) & 2.23(7) &  0.394(12) & 0.198(8) & 0.0481(19) & 0.0120(9) \\
$m$ & 1.837(18) &   3.110(42) &  3.368(23) &  4.070(30)  & 4.086(21) & 4.260(25) & 4.165(40) \\
\hline
$\tilde{\chi^2}$ & 4.4 & 9.4  & 15.7   &  8.9 &  5.7 &    6.0  &      15.7 \\
$\eta$         &   5.0~\mbox{$10^{-4}$}   &  5.4~\mbox{$10^{-4}$}    &
 3.4~\mbox{$10^{-4}$}    &    1.1~\mbox{$10^{-4}$}   &    5.7~\mbox{$10^{-5}$}
 &  5.4~\mbox{$10^{-5}$}   &   4.4~\mbox{$10^{-5}$}  \\ 
\end{tabular}
\end{ruledtabular}
}
\end{table*}

\begin{figure}[!ht]
\centering
\includegraphics[width=\columnwidth]{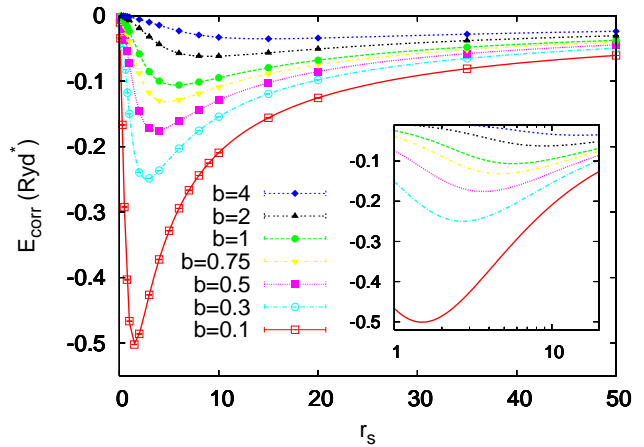}
\caption{(Color online) Correlation energy versus the density
  parameter $r_s$. The points are the LRDMC data used in the fit. Their
  dimension is much bigger than the error bars. We computed the LRDMC
  correlation energy at
  $r_s=0.1,0.2,0.3,0.4,0.6,0.8,1,2,3,4,6,8,10,15,20,35,50$.
  The curves are the
  parametrization of the correlation energy for various values of the width parameter $b$.
  In the inset, the semi-log plot magnifies the region around the minimum of
  the energy correlation functional.} 
\label{plot_corr}
\end{figure}

We define the accuracy $\eta$ of the parametrization by
\begin{equation}
\label{fit_accuracy}
\eta=\frac{1}{M} \sum_{i=1}^M |E_{corr}(i) - \epsilon(i)|,   
\end{equation}
where $M$ is the total number of $r_s$ points computed for a  given $b$. In
practice, $\eta$ is the average of the residuals and measures the discrepancy
between the computed value $E_{corr}$ and its
parametric value $\epsilon$. The order of magnitude of the accuracy $\eta$ 
is between $10^{-4} Ryd^*$ and $10^{-5} Ryd^*$, depending on the thickness of
the wire.  
It means that the parametrization is very accurate in the range of density
with $0<r_s<50$ and for values $0.1 \le b \le 4$ of the width parameter.
In Fig.~\ref{plot_corr} we plot the points and the parametrization curves for
the correlation energy in the range of density and width taken into account. 
Notice that the thiner is the wire, the more correlated is its ground state.

\begin{figure}[!ht]
\centering
\includegraphics[width=\columnwidth]{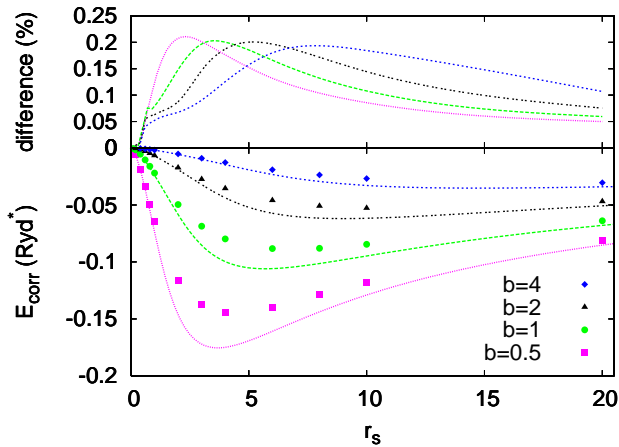}
\caption{(Color online) In the lower panel, we plot the correlation energy obtained by
  Calmels and Gold~\cite{calmels} (points) and the parametrization of our LRDMC
  data (curves) for various values of the width parameter $b$. In the upper
  panel, we compute the relative difference $|\epsilon-
  E_{Calmels}|/\epsilon$ interpolated all over the density range taken into account.}
\label{plot_gold3}
\end{figure}

To conclude this section, in Fig.~\ref{plot_gold3} 
we compare our LRDMC results with those obtained by Calmels and Gold using the 3SRA
approach\cite{calmels}. 
The relative difference between the two methods is about $20\%$ around the minimum, where
the absolute value of the correlation energy is larger. 
The STLS method with the 3SRA sum rule is able to yield a large fraction
($80\%$ or more) of the correlation energy. 
However our results, which are formally exact,
represent a further improvement with respect to the previous estimate of the
correlation energy for this model quantum wire.

\section{Pair correlations}
\label{pair_correlations}

\begin{figure}[!ht]
\centering
\includegraphics[width=\columnwidth]{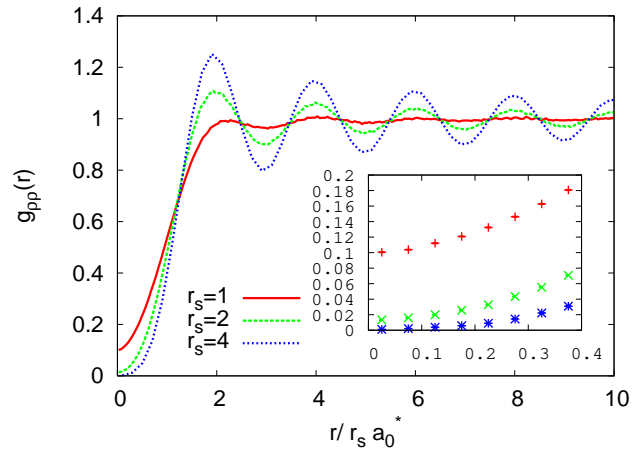}
\caption{(Color online) Charge-charge pair correlation function 
$g_{\rho\rho}(r)$ computed from LRDMC simulations with $b=0.1$ and $N=82$ at
  different densities. In the inset, the short-range part of $g_{\rho\rho}(r)$
is magnified.}
\label{g_nn}
\end{figure}

\begin{figure}[!ht]
\centering
\includegraphics[width=\columnwidth]{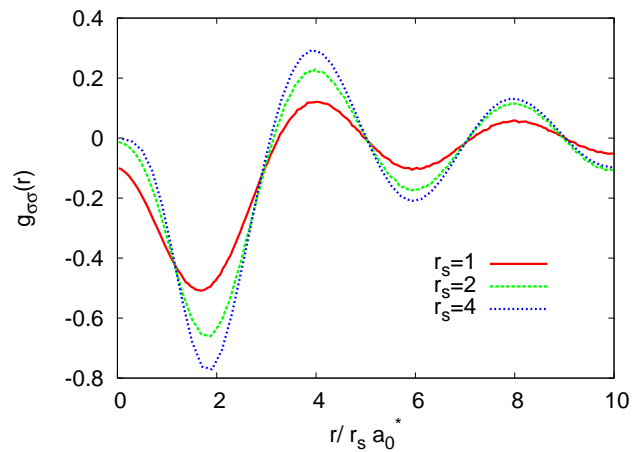}
\caption{(Color online) Spin-spin pair correlation function 
$g_{\sigma\sigma}(r)$ computed from LRDMC simulations with $b=0.1$ and $N=82$.}
\label{g_mm}
\end{figure}

In this section we study 
the density-density and spin-spin correlation functions, which are useful 
quantities for assessing the nature of the ground state of the system.
We first analyze the charge $g_{\rho\rho}(r)$ and spin $g_{\sigma\sigma}(r)$
pair correlations, defined as
\begin{eqnarray}
g_{\rho\rho}(r) & = & 1/2 \left( g_{\uparrow \uparrow}(r) + g_{\uparrow
  \downarrow}(r) \right),  \\ \nonumber
g_{\sigma\sigma}(r) & = & 1/2 \left( g_{\uparrow \uparrow}(r) - g_{\uparrow
  \downarrow}(r) \right),  
\end{eqnarray}
where we used the spin resolved pair distribution function
\begin{equation}
g_{\alpha \beta}(r) = \frac{1}{L \rho_\alpha \rho_\beta} \sum_{i \ne j} \langle
\delta(r_i^\alpha -  r_j^\beta -r) \rangle, 
\end{equation}
with $\alpha,\beta=\uparrow,\downarrow$, and $\rho_\uparrow=\rho_\downarrow=\rho/2$
is the density of the two spin components in the unpolarized system.
We computed $g_{\rho\rho}(r)$ and $g_{\sigma\sigma}(r)$ of the wires with
thickness $b=0.1$ and densities $r_s=1,2,4$, by means of the LRDMC algorithm and the
forward walking technique\cite{calandra}, which yields unbiased expectation values
on the ground state of the system also for operators which do not commute with
the Hamiltonian. The correlation functions are drawn in Figs.~\ref{g_nn} and \ref{g_mm}.

The charge-charge correlation functions in Fig.~\ref{g_nn} reveal the strong effect
of the electronic correlation in the low density regime. 
As the density is decreased, the particles repel each other with an effective interaction
which is stronger. Consequently $g_{\rho\rho}(0)$ is
smaller, as one can see in the inset of the Fig.~\ref{g_nn}. At the same time
the fluctuations in the $g_{\rho\rho}(r)$ are larger, with periodicity $2 r_s
a_0^*$, and a slow decay. We will see however that this decay is not slow enough to give
rise to a true long range order (Wigner crystal), since the quantum
fluctuations will prevent the freezing of the charge in 1D.
On the other hand, the short-range behavior of the spin-spin correlation
functions in Fig.~\ref{g_mm} shows the antiferromagnetic character of the coupling 
among electrons, and the underlying periodicity in the spin sector is $4
r_s a_0^*$, i.e. twice the mean interparticle distance. 

The analytical behavior of charge and spin correlations for the quasi 1 DEG
with long range interaction has been obtained by Schulz\cite{schulz} using bosonization
techniques applied to an effective one dimensional Hamiltonian with  
linearized kinetic energy. For that model Hamiltonian, it turns out that the
charge correlation function exhibits a slow decay of its $4 k_F$
component
\begin{equation} 
\langle \rho(x) \rho(0) \rangle  \simeq A \cos(4 k_F x) \exp(-4 c \sqrt{\ln(x)}),
\label{charge}
\end{equation}
with $c$ an interaction dependent parameter. Its behavior has been related 
to a quasi order of the electrons, 
since in one dimension there is no true long-range order.
Their fluctuations are almost frozen to
maximize the interparticle distance along the wire, 
and their relative positions are pinned
around lattice sites with periodicity $2 r_s a_0^*$, exactly as we have seen
in the $g_{\sigma\sigma}(r)$ for our model. 
The formation of this quasi Wigner
crystal comes from the $1/r$ tail of the potential, since in 
the case of short-range interactions the $4 k_F$ component of the charge 
correlation function decays much faster.
On the other hand the spin correlation function has no singularity at $4 k_F$
and exhibits the slowest decay for the $2 k_F$ component. Indeed, 
according to the bosonization technique\cite{schulz},  the large 
distance spin  correlations are given by:
\begin{equation}
\langle \sigma(x) \sigma(0) \rangle \simeq B \cos(2 k_F x) \exp(-c \sqrt{\ln(x)})/x,
\label{spin}
\end{equation}
where $c$ is the same as in Eq.~\ref{charge}.

\begin{figure}[!ht]
\centering
\includegraphics[width=\columnwidth]{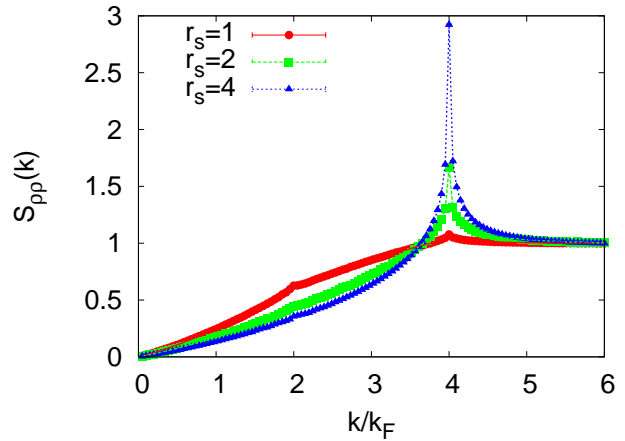}
\caption{(Color online) Charge structure factor $S_{\rho\rho}(k)$ computed from LRDMC 
simulations with $b=0.1$ and $N=82$.}
\label{chargestr_full}
\end{figure}

\begin{figure}[!ht]
\centering
\includegraphics[width=\columnwidth]{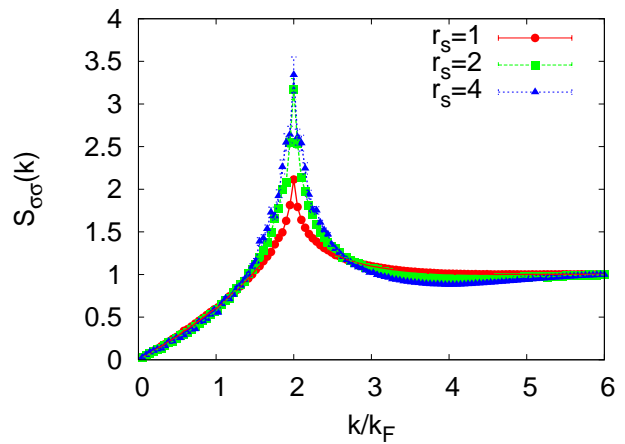}
\caption{(Color online) Spin structure factor $S_{\sigma\sigma}(k)$ computed from LRDMC 
simulations with $b=0.1$ and $N=82$.}
\label{spinstr_full}
\end{figure}

In order to check whether the Eqs.~\ref{charge} and~\ref{spin} are valid
for our quantum wire model with long-range interactions and
quadratic dispersion, we have also studied the charge $S_{\rho\rho}(k)$
and spin $S_{\sigma\sigma}(k)$ structure factors, defined as
\begin{eqnarray}
S_{\rho\rho}(k) & = & \langle \rho(k) \rho(-k) \rangle/N \\ \nonumber
S_{\sigma\sigma}(k) & = & \langle \sigma(k) \sigma(-k) \rangle/N,
\end{eqnarray}
where $\rho(k)$ ($\sigma(k)$) is the Fourier component of the local charge
(spin) density. We have computed 
the structure factors at several $k$ values
for $b=0.1$, $r_s=1,2,4$, using $N=10,22,42,62,82$ in the LRDMC calculations
and $N=10,22,42,82,162$ in the less expensive VMC calculations.
In Figs.~\ref{chargestr_full} and \ref{spinstr_full} we plot $S_{\rho\rho}(k)$
and  $S_{\sigma\sigma}(k)$ for the LRDMC simulations with the largest number
of particles ($N=82$). 
While the spin-spin correlations show a peak at $k=2 k_F$, 
for the charge degrees of freedom the highest peak arises 
at $k=4 k_F$, which corresponds to the periodicity 
of the quasi Wigner crystal.  

\begin{figure}[!ht]
\centering
\includegraphics[width=\columnwidth]{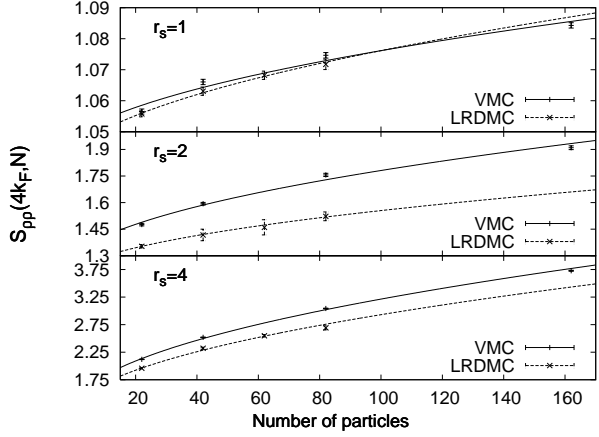}
\caption{Peak of the charge structure factor at $4 k_F$ versus the system size
for $r_s=1,2,4$, $b=0.1$, obtained from VMC and LRDMC calculations. The curves fit
the points with function in Eq.~\ref{charge_fit}.}
\label{peaknum}
\end{figure}

\begin{figure}[!ht]
\centering
\includegraphics[width=\columnwidth]{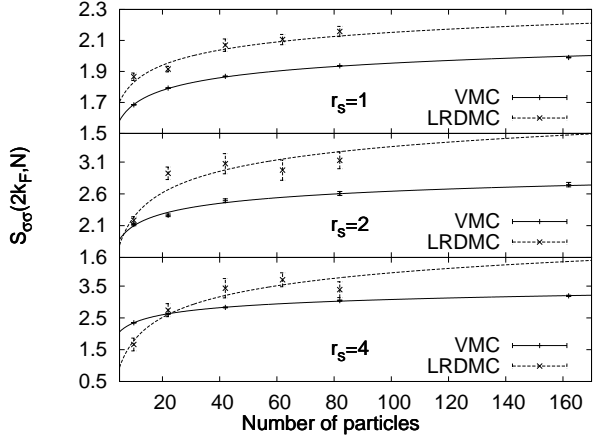}
\caption{Peak of the spin structure factor at $2 k_F$ versus the system size
for $r_s=1,2,4$, $b=0.1$, obtained from VMC and LRDMC calculations. The curves fit
the points with function in Eq.~\ref{spin_fit}.}
\label{peakspin}
\end{figure}

In particular, we have studied the dependence 
of the peak heights $S_{\rho\rho}(4k_F,N)$ and $S_{\sigma\sigma}(2k_F,N)$  
on the number of particles $N$, and compared it with the behavior predicted by
the bosonization. From Eqs. \ref{charge} and \ref{spin}, one can easily
obtain\cite{comment}: 
\begin{eqnarray}
S_{\rho\rho}(4k_F,N) & = & a_1 L \exp (-4 c \sqrt{\log{L}}) + a_2, 
\label{charge_fit}\\
S_{\sigma\sigma}(2k_F,N) & = & 
a_3 ( \sqrt{\log{L}}/c + 1/c^2 ) \exp (- c\sqrt{\log{L}}) \nonumber \\
&+& a_4,
\label{spin_fit}
\end{eqnarray}
with $L=2r_sN$, $a_1$,$a_2$,$a_3$,$a_4$ model and density dependent
parameters, and $c$ the same as in the bosonization results. 
We have then used Eqs. \ref{spin_fit} and \ref{charge_fit} 
to fit our results, obtaining first $a_1,a_2,c$
from the fit of $S_{\rho\rho}(4 k_F,N)$ and then $a_3,a_4$ from
$S_{\sigma\sigma}(2 k_F,N)$, with $c$ fixed by $S_{\rho\rho}(4 k_F,N)$.
In Figs.~\ref{peaknum} and \ref{peakspin} we plot
the curves which best interpolate the VMC and LRDMC values for 
 $S_{\rho\rho}(4 k_F,N)$ and $S_{\sigma\sigma}(2 k_F,N)$.

Our VMC and LRDMC results seem consistent with the predictions of the
bosonization, at least in the range of system sizes taken into account. 
As it is apparent from Figs.~\ref{peaknum} and \ref{peakspin}, the points are
well fitted by Eqs.~\ref{charge_fit} and \ref{spin_fit} respectively, the
distance from the interpolating curves being usually less than 2 standard deviations.
As already reported in Ref.~\onlinecite{mc}, VMC results 
are in qualitative agreement with the bosonization findings, and the
variational ansatz in Eq.~\ref{wf} 
is good enough to capture the correct ground state properties.
Indeed the LRDMC projection changes only quantitatively the VMC points, by
reducing the charge structure factor, 
and by enhancing the spin structure factor, which however remains 
finite in the thermodynamic limit in accordance with Eq.~\ref{spin_fit}.
We emphasize once  again   that for this one dimensional system
the LRDMC results are ``exact'' within their statistical accuracy, since the
LRDMC yields the exact ground state energy \emph{and} 
the points in Fig~\ref{peaknum} and \ref{peakspin} are obtained
using the forward walking technique\cite{calandra}, 
which provides an unbiased expectation value for each correlation function.

\begin{figure}[!ht]
\centering
\includegraphics[width=\columnwidth]{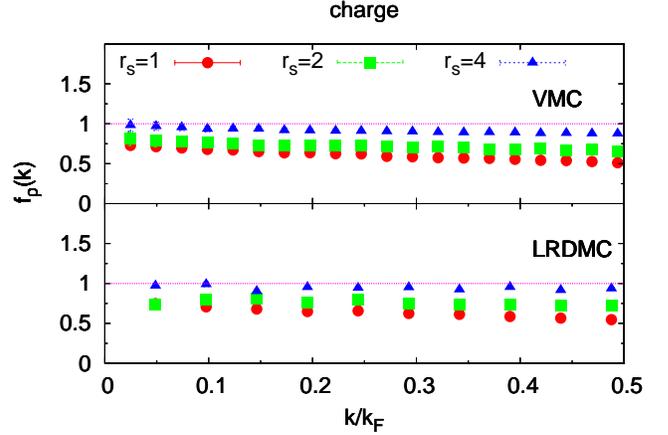}
\caption{(Color online) Plot of $f_\rho(k)=\frac{2}{\sqrt{r_s}} \frac{\sqrt{|\ln(k)|}}{k}
  S_{\rho\rho}(k)$ versus $k/k_F$ for $r_s=1,2,4$, $b=0.1$, obtained from VMC ($N=162$)
  and DMC ($N=82$) calculations. The horizontal line is drawn as an eye guide.} 
\label{chargespec}
\end{figure}

\begin{figure}[!ht]
\centering
\includegraphics[width=\columnwidth]{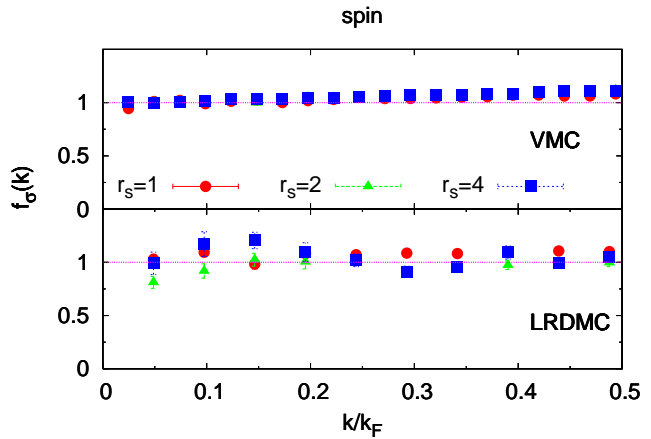}
\caption{(Color online) Plot of $f_\sigma(k)=\frac{v_F}{k} S_{\sigma\sigma}(k)$ 
  versus $k/k_F$ for $r_s=1,2,4$, $b=0.1$, obtained from VMC ($N=162$)
  and DMC ($N=82$) calculations. The horizontal line is drawn as an eye
  guide.}
\label{spinspec}
\end{figure}

The behavior of the spin and charge structure factor at small momenta
reveals important features of the ground state, which are related to the
low energy modes of the system. According to the bosonization results~\cite{parola},
$S_{\rho\rho}(k)$ should behave as $\propto |k|/\sqrt{|\ln(k)|}$, while $S_{\sigma\sigma}(k)$
should go linearly with $|k|$. Our results are plotted in Figs.~\ref{chargespec}
and \ref{spinspec}. 
In particular, in Fig.~\ref{chargespec} we draw the
renormalized charge structure factor $f_\rho(k)=\frac{2}{\sqrt{r_s}}
\frac{\sqrt{|\ln(k)|}}{k} S_{\rho\rho}(k)$. 
Both the VMC and LRDMC data are in agreement with the small $k$ limit of
$S_{\rho\rho}(k)$:
\begin{equation}
\lim_{k \rightarrow 0} S_{\rho\rho}(k) = \alpha_g \frac{\sqrt{r_s}}{2} \frac{|k|}{\sqrt{|\ln(k)|}},
\label{s_small_k}
\end{equation}
where $\alpha_g$ is a factor which seems close to 1 and only slightly dependent
on $r_s$, although the logarithmic behavior of $f_\rho(k)$ would require
much smaller $k$ in Fig.~\ref{chargespec} in order to have an accurate extrapolation 
for $\alpha_g$. Anyway, our result agrees at least qualitatively 
with the bosonization findings. Moreover, it is possible to derive analytically 
the behavior in Eq.~\ref{s_small_k} directly from our variational wave function $\Psi_T$, 
using the expression obtained by Reatto and Chester \cite{reatto}, 
that relates the small momenta behavior of $S_{\rho\rho}(k)$ 
with the 2-body Jastrow factor included in $\Psi_T$:
\begin{equation}
\label{reatto_rel}
S_{\rho\rho}(k) \simeq \frac{S_0(k)}{1 + 2 \rho u(k) S_0(k)},
\end{equation}
where $u(k)$ is the Fourier transform of the Jastrow function.
The above relation is approximate for finite $k$, but it
becomes exact to the leading order in $k$, for $k \rightarrow 0$, for
the variational structure factor, i.e., for the one evaluated from the
Slater-Jastrow trial function. After plugging the definition 
of $u(k)$ (Eqs.~\ref{u_rpa} and
\ref{u_rpa_opt}) into the above equation, 
in the limit $k \rightarrow 0$ we have
\begin{equation}
\label{s_reatto}
S_{\rho\rho}(k) = \frac{1}{\gamma} \frac{\sqrt{r_s}}{2}
\frac{|k|}{\sqrt{|\ln(k)|}},  
\end{equation}
where $\gamma$ is the parameter of the optimized RPA Jastrow factor. It is
therefore clear that our choice of $u(k)$ satisfies the
correct behavior of the charge structure factor already at the variational
level, and the LRDMC simulations do not change this behavior (see
Fig.~\ref{chargespec}).
Moreover, from the knowledge of $S_{\rho\rho}(k)$ we can infer the
behavior of the low energy charge excitations (plasmons) $\omega_\rho(k)$. 
Indeed, a variational estimate of $\omega_\rho(k)$ is given by (see Appendix
\ref{appendix_fsum})
\begin{equation}
\omega_\rho(k)=\frac{k^2}{S_{\rho \rho}(k)},
\end{equation}
and in the limit $k \rightarrow 0$ it turns out that
\begin{equation}
\label{our_limit}
\omega_\rho(k)=\frac{2}{\alpha_g~\sqrt{r_s}} |k| \sqrt{|\ln k|}.
\end{equation}
This expression should be compared with the low energy spectrum provided by
bosonization studies\cite{schulz,fabrizio,wang} of the Coulomb Luttinger
liquid, which reads
\begin{eqnarray}
  \omega_\rho(k) & = & v_\rho(k) |k|, \nonumber \\
v_\rho(k) & = & v_F \sqrt{(1 + g_1) (1 - g_1 + 2 V(k)/ \pi v_F)},
\end{eqnarray}
where $v_\rho$ is the charge velocity, 
and $g_1$ is the amplitude of the backward scattering process. For small $k$
excitations, we have
\begin{equation}
\label{charge_excitation}
\lim_{k \rightarrow 0} \omega_\rho(k) = \sqrt{1+g_1}\frac{2}{\sqrt{r_s}} |k|
\sqrt{|\ln k|},
\end{equation}
which corresponds to our findings in Eq.~\ref{our_limit}.
Since in one dimension the long-wavelength spin and charge modes are
independent, they have different velocities. 
This difference, due to the so-called spin-charge separation, has been seen in
a remarkable experiment by Auslaender \emph{et al.} \cite{auslaender05}, and
predicted by the Luttinger liquid theory. Indeed, according to this theory 
the spin excitations are
\begin{equation}
\omega_\sigma(k) = v_F \sqrt{1 - g_1^2} |k|.
\end{equation}
The spin dispersion is linear, since it is not affected by the long-range tail
of the Coulomb interaction, in contrast to the charge velocity $v_\rho$, which
is renormalized by the Fourier transform of the potential at small $k$. The
linear behavior of the spin branch is reflected by the linear decay of 
$S_{\sigma\sigma}(k)$ as $k$ goes to 0. In
Fig.~\ref{spinspec} we plot the renormalized spin structure factor 
$f_\sigma(k)=\frac{v_F}{k} S_{\sigma\sigma}(k)$ computed by carrying out both
VMC and LRDMC simulations for $r_s=1,2,4$. Since the value of $f_\sigma(k)$ is 1 in
the limit of small $k$, it turns out that
\begin{equation}
S_{\sigma\sigma}(k) = \frac{k}{v_F},
\end{equation}
and it is the same behavior as the spin structure factor of a non-interacting
gas. Therefore, the interaction leaves unchanged the  $S_{\sigma\sigma}(k)$  tail at
small $k$, which is another striking feature of the spin-charge separation.
Notice that if we use $S_{\sigma\sigma}(k)$ to estimate the low lying spin-wave
excitations (see Appendix \ref{appendix_fsum}), we will obtain a spin velocity
equal to $v_F$, and independent of $r_s$. This is of course a variational
estimate, since the true spin velocity strongly depends on the density
and is significantly reduced by the effective interaction\cite{macdonald}, 
being equal to $v_F$ only in the high density weak interaction regime.

\section{Conclusions}
\label{conclusions}
In this paper we have carried out extensive Monte Carlo simulations to compute
the ground state properties of the quantum wire model with unscreened long
range interactions. We have used the novel LRDMC framework, which is shown to
be more efficient than the standard DMC algorithm in the strong coupling
regime, i.e. at low densities, when the exchange is extremely small and the
features of a quasi-Wigner crystal are manifest. We computed the exact
correlation energy, and found a simple and accurate parametrization, which
fits the correlation energy over a wide range of electron densities and
lateral confinements. This parametrization includes the correct behavior at
high densities ($\epsilon(r_s \rightarrow 0) \propto r_s^2$), given by the RPA
approximation. On the other hand, we guessed the asymptotic behavior 
at low densities from our data, and we found that $\epsilon(r_s \rightarrow
\infty) \propto -\ln(r_s)/r_s$ fits well the LRDMC correlation energies in the
strong coupling regime. 
We believe that our parametrization provides an extremely reliable functional 
for further DFT computations of quasi one dimensional systems. 
Last but not least, we showed that the pair
correlations of our model, exactly computed by means of LRDMC simulations and
forward walking techniques, reproduce all features of the so-called Coulomb
Luttinger liquid, i.e. a Luttinger liquid with linear dispersion and 
long-range Coulomb interaction. In particular, our results are compatible with 
the slow decay ($\propto \exp(-4c\sqrt{\ln(x)})$ of the spin structure factor
at $4 k_F$, which is the signature of a quasi order of the charge degrees of
freedom. Moreover, the small $k$ behavior of the static structure
factor is in agreement with the bosonization findings, both for the charge and
spin modes.

We plan to extend this study also to the spin polarized case, and to provide a
spin resolved energy functional to be used as input of one dimensional DFT
calculations with local spin density approximation.

\acknowledgments 
We thank M. Fabrizio, C. J. Umrigar, D. M. Ceperley, S. Moroni
and M. R. Geller for useful discussions.
This work was partially supported by COFIN 2005, and CNR. One of us (M.C.)
acknowledges support in the form of the NSF grant DMR-0404853.

\appendix
\section{RPA calculation of the correlation energy}
\label{appendix_RPA}
In this appendix, we compute the correlation energy in the high density limit
using the random phase approximation (RPA). The RPA correlation energy reads
\cite{pines}
\begin{eqnarray}
\label{RPA_corr}
E^{RPA}_{corr} & = & \frac{L}{2 \pi} \int^{+\infty}_{-\infty} dk~E(k),
\nonumber \\
E(k) & = & \frac{1}{4 \pi} \frac{|k|}{N} \int^{+\infty}_{-\infty} d\lambda ~
\ln( 1 - v(kb) \chi^0(k,ik\lambda) ) \nonumber \\
 & & + v(kb) \chi^0(k,ik\lambda), 
\end{eqnarray}
where $v(kb)$ is the Fourier transform of the potential, 
\begin{equation}
\label{response}
\chi^0(k,\omega)=\frac{1}{2 \pi k} \ln\left( \frac{\omega^2-(k^2-v_F
  k)^2}{\omega^2 - (k^2+v_F k)^2} \right), 
\end{equation}
and the physical density-density response function of the free 1D
electron gas is \begin{equation}\lim_{\eta\rightarrow 0^+}\chi^0(k,\omega+i\eta).\end{equation}
Following the seminal work of Gell-Mann and Brueckner \cite{gellmann}, whose
approach has also been used by Rajagopal and Kimball \cite{rajagopal} for the
2D case, we define the so called electron-hole propagator
\begin{eqnarray}
Q_q(u) & = & \int^{+\infty}_{-\infty} dk ~ \int^{+\infty}_{-\infty} dt ~
f(k) (1-f(k+q)) \nonumber \\
& & e^{-ituq} \exp(-|t| (\frac{1}{2} q^2 + kq)) , 
\end{eqnarray}
where $f(x)=\theta(|x|-1)$ is the zero temperature Fermi distribution, 
with $\theta$ the step function:
\begin{equation}
\theta(x) = \left\{ 
\begin{array}{ll}
1 & \textrm{if $x<0$}\\
0 & \textrm{if $x \ge 0$}.
\end{array} \right. 
\end{equation}
It is easy to see that the response function in Eq.~\ref{response} can be
simply related to the electron-hole propagator by the relation
\begin{equation}
\chi^0(k,\omega)=-\frac{1}{2 \pi k_F} Q_q(u),
\end{equation}
where $q$ and $u$ are dimensionless  variables defined by
\begin{eqnarray}
k & = & k_F q \nonumber \\
\omega & = & i~k_F q ~ v_F u.
\end{eqnarray}
Therefore, if we rewrite Eq.~\ref{RPA_corr} in terms of the electron-hole
propagator and using the dimensionless variables $u$ and $q$, we obtain
\begin{widetext}
\begin{eqnarray}
\label{RPA_corr2}
E^{RPA}_{corr}(r_s) & = & \frac{1}{2 \pi (\alpha r_s)^2} \int^{+\infty}_0 dq~
\int^{+\infty}_0 du ~ q \left\{ \ln \left(1 + \frac{\alpha r_s}{2 \pi}
~v\left(\frac{qb}{\alpha r_s}\right) Q_q(u)\right) - 
\frac{\alpha r_s}{2 \pi} ~v\left(\frac{qb}{\alpha
  r_s}\right) Q_q(u) \right\} \nonumber \\
  & = &  \frac{1}{2 \pi (\alpha r_s)^2} \int^{+\infty}_0 dq~
\int^{+\infty}_0 du ~ q \sum_{n=2}^{+\infty} \frac{(-1)^{n-1}}{n}
\left(\frac{\alpha r_s}{2 \pi} \right)^n ~v^n\left(\frac{qb}{\alpha
  r_s}\right) Q^n_q(u), 
\end{eqnarray}
\end{widetext}
where $\alpha = 4 / \pi$ in 1D.  
To pass from the first line to the second one we have Taylor expanded the logarithm.
Notice that in 1D, in contrast to the 3D
and 2D cases, the integrals converge in \emph{all} orders of the expansion,
since $v(qb)$ ($=\tilde{V}_b(q)$ in Eq.~\ref{Fourier_potential}) 
diverges only logarithmically at small $q$:
\begin{equation}
\label{potential_asymp}
v(qb) \simeq \left\{ 
\begin{array} {ll}
-4 \ln(qb) & \textrm{if $q \rightarrow 0$} \\
\frac{2}{q^2 b^2} & \textrm{if $q \rightarrow +\infty$}.
\end{array} \right. 
\end{equation}
Moreover, the leading order in the $r_s$ expansion for Eq.~\ref{RPA_corr2} is
given by $n=2$ (i.e. the direct lowest ring diagram in the perturbative
expansion of the interaction). Since we are interested in the lowest order
$r_s$ expansion of the RPA correlation energy, we keep the term with $n=2$ and
we discard the others. Thus, we get
\begin{equation}
\label{RPA_approx}
E^{RPA}_{corr} \simeq -\frac{1}{2} \frac{1}{(2 \pi)^3} \int^{+\infty}_0 dq~
 q ~ v^2\left(\frac{qb}{\alpha r_s}\right) F(q),
\end{equation}
with $F(q)= \int^{+\infty}_0 du ~ Q^2_q(u)$ the integral over the
dimensionless frequency of the 2-particles 
electron-hole propagator at a given momentum
transfer $q$. Indeed $F(q)$ can be explicitly written as
\begin{equation}
\label{F(q)}
F(q)=\frac{2 \pi}{q} \int_{1-q}^1 dk_1 \int_{1-q}^1 dk_2
~\frac{1}{q^2+q(k_1+k_2)},
\end{equation}
and for zero $q$-transfer $F(0)=\pi$. If we rescale the variable $q$ in the
integration \ref{RPA_approx} ($q \rightarrow \frac{\alpha r_s}{b} q$), and we
keep the lowest order in $r_s$, the RPA correlation energy reads:
\begin{equation}
E^{RPA}_{corr} = -\frac{1}{(4 \pi)^2} \left(\frac{\alpha
  r_s}{b}\right)^2 \int_0^{+\infty} dz~ z~ v^2(z).
\end{equation}  
In order to make sure that this is the correct high energy limit of the
correlation energy, we need to consider also the second order exchange
contribution in the perturbation theory. It is neglected in the RPA
approximation, but can yield non trivial corrections in the
$r_s$ expansion, like in the two~\cite{rajagopal} and three~\cite{gellmann}
dimensional electron gas. The second order exchange is
\begin{equation}
E^{II}_{exch} = \frac{1}{4} \frac{1}{(2 \pi)^3} \int_0^{+\infty}
dq~q~v\left(\frac{qb}{\alpha r_s}\right) F_{exch}(q),
\end{equation}
where now
\begin{eqnarray}
F_{exch}(q) & = &   \frac{2 \pi}{q} \int_{1-q}^1 dk_1 \int_{1-q}^1 dk_2  
\nonumber \\
& & v\left(\frac{(q+k_1+k_2)b}{\alpha r_s} \right) \frac{1}{q^2+q(k_1+k_2)}.
\end{eqnarray}
One can easily see that $F_{exch}(0)=\pi~ v\left(\frac{2 b}{\alpha r_s} \right)$.
The difference from the second order direct ring (Eqs.~\ref{RPA_approx}
and \ref{F(q)}) is the vertex interaction $V(q)$ computed at $q+k_1+k_2$
instead of $q$, and the overall factor is reduced by a factor of 2, due to the
spin summation which is now restricted only to the parallel contribution.
If we rescale the variable $q$ as before ($q \rightarrow \frac{\alpha r_s}{b}
q$), we find
\begin{eqnarray}
 E^{II}_{exch} & = & \frac{1}{2} \frac{1}{(4 \pi)^2} \left(\frac{\alpha
  r_s}{b}\right)^2 \int_0^{+\infty} dz~ z~ v(z)~ v\left(\frac{2 b}{\alpha r_s}
  \right)
\nonumber \\
& = & \frac{4 A_{exch}}{\pi^6 b^4} r_s^4,
\end{eqnarray}  
where we used the asymptotic behavior of the potential $v(x)$
for $x \rightarrow +\infty$ (Eq.~\ref{potential_asymp}), and defined
$A_{exch}=\int_0^{+\infty} dz~ z~ v(z)$. It is apparent that the exchange second
order diagram contributes only to the fourth order of $r_s$. 
Therefore, at the lowest order in $r_s$, the correlation energy is:
\begin{equation}
E_{corr} = - \frac{A}{\pi^4 b^2} r_s^2,  
\end{equation}
with $A=\int_0^{+\infty} dz~ z~ v^2(z)= 4.9348$. This result turns out to be the
same as a high density extrapolation\cite{calmels} of the correlation energy
for the same model studied here, 
obtained by Gold and Calmels within the so
called mean spherical approximation (MSA). In two and three dimensions, the MSA
yielded high density expressions of the correlation energy which  
were slightly different from the RPA findings\cite{gold}. 
In this case however, the MSA
and the RPA results are in perfect agreement.

\section{Variational energies of charge excitations}
\label{appendix_fsum}
Let $|\Psi_0\rangle$ be the ground state of the Hamiltonian
\begin{equation}
\label{ham_here}
H= \sum_{k,\sigma} \epsilon(k) c^\dag_{k,\sigma} c_{k,\sigma} + \frac{1}{2L}
 \sum_{q \ne 0}  V(q) (\rho_{-q} \rho_q - N),
\end{equation}
where $\epsilon(k)$ is the dispersion of the non interacting system, $V(q)$ is
the Fourier transform of the interaction, and $\rho_q = \sum_{k,\sigma} c^\dag_{k+q,\sigma}
c_{k,\sigma}$ is the Fourier transform of the charge density operator. 
In analogy with the Feynman's construction for the liquid Helium\cite{feynman},
a variational wave function for a charge excitation (plasmon) with momentum $q$ is given by
\begin{equation}
|\Psi_q\rangle = \rho_q |\Psi_0\rangle.
\end{equation}
Its variational energy $\langle \Psi_q | H | \Psi_q\rangle / \langle \Psi_q |
\Psi_q\rangle$ is $E_q$, while the GS energy is $E_0$. Notice that the
normalization of $|\Psi_q\rangle$ is
\begin{equation}
\langle \Psi_q | \Psi_q \rangle = N S_{\rho\rho}(q) \langle \Psi_0 | \Psi_0\rangle,
\end{equation}
where $S_{\rho \rho}(q)=\frac{1}{N} \langle \rho_{-q} \rho_q \rangle$ is the static charge
structure factor.
We are now going to find an expression which allows us to estimate
the excitation energy of a plasmon with momentum $q$ from knowledge of
$S_{\rho \rho} (q)$.
We start by evaluating the double commutator
\begin{equation}
\label{double_comm_1}
[[\rho_q,H],\rho_{-q}]=\sum_k (\epsilon_{k+q}+\epsilon_{k-q}-2\epsilon_k) c^\dag_k c_k,
\end{equation}
where we used the Hamiltonian in Eq.~\ref{ham_here}. On the other hand, we have
\begin{equation}
\label{double_comm_2}
\frac{\langle \Psi_0 | [[\rho_q,H],\rho_{-q}] | \Psi_0 \rangle}{\langle \Psi_0 |
  \Psi_0 \rangle}=2 N S_{\rho \rho}(q) (E_q - E_0),  
\end{equation}
by applying the definition of the double commutator $[[\rho_q,H],\rho_{-q}]=\rho_q H
\rho_{-q} + \rho_{-q} H \rho_q - \rho_q \rho_{-q} H - H \rho_q \rho_{-q}$. Merging
Eq.~\ref{double_comm_1} and Eq.~\ref{double_comm_2}, we are led to the
following identity:
\begin{equation}
\label{fsum}
E_q - E_0 = \frac{ \langle \Psi_0 | \sum_k
  (\epsilon_{k+q}+\epsilon_{k-q}-2\epsilon_k) c^\dag_k c_k | \Psi_0
  \rangle }{2 N S_{\rho \rho}(q) \langle \Psi_0 | \Psi_0 \rangle}.
\end{equation}
$E_q-E_0$ is an estimate of the plasmon excitation with momentum
$q$. In general the ansatz $|\Psi_q\rangle = \rho_q |\Psi_0\rangle$ 
is not exact for the lowest energy wave function with momentum
$q$, but gives a variational energy, since it belongs to
the same $q$-subspace as the true excited state and 
is orthogonal to subspaces with different $q^\prime$.
In the limit of $q$ small, Eq.~\ref{fsum} turns out to be
\begin{eqnarray}
\label{fsum2}
E_q - E_0 & \simeq & 
\frac{\langle \Psi_0 | \sum_k \partial^2_k \epsilon_k
  c^\dag_k c_k | \Psi_0 \rangle}{2 N \langle \Psi_0 | \Psi_0 \rangle} 
\frac{q^2}{S_{\rho \rho}(q)}
\nonumber \\
&\simeq & \frac{q^2}{S_{\rho \rho}(q)}, 
\end{eqnarray}
where we used the quadratic dispersion $\epsilon(k)=k^2$ in $Ryd^*$ units.
Therefore, from the knowledge of $S_{\rho \rho}(k)$ calculated for the GS of the
system we can evaluate its excitation spectrum in a variational way. 
Moreover the smaller $q$ is, the better Eq.~\ref{fsum2} approximates  
the true plasmon energy.
In the same way, one can estimate the energy of the spin-wave 
excitations (spinons), by using the relation in Eq.~\ref{fsum2} with 
$S_{\rho \rho}(k)$ replaced by the static spin structure factor 
$S_{\sigma \sigma}(k)$.

\end{document}